\documentclass[aps,twocolumn,prc,epsfig,showpacs]{revtex4}
\usepackage{graphicx}
\usepackage{epsfig}
\begin{document}
\title{Scaling predictions for radii of weakly bound triatomic
molecules}
\author{M. T. Yamashita}
\address
{Laborat\'orio do Acelerador Linear, Instituto de
F\'{\i}sica, Universidade de S\~{a}o Paulo, C.P. 66318, CEP
05315-970, S\~{a}o Paulo, Brazil}
\author{R. S. Marques de Carvalho, Lauro Tomio}
\address
{Instituto de F\'\i sica Te\'orica, Universidade
Estadual Paulista, 01405-900, S\~{a}o Paulo, Brazil}
\author{T. Frederico}
\address
{Departamento de F\'\i sica, Instituto Tecnol\'ogico de
Aeron\'autica, Centro T\'ecnico Aeroespacial, 12228-900, S\~ao
Jos\'e dos Campos, Brazil}

\date{\today}

\begin{abstract}
The mean-square radii of the molecules $^4$He$_3$, $^4$He$_2-^6$Li,
$^4$He$_2-^7$Li and $^4$He$_2-^{23}$Na are calculated using a three-body
model with contact interactions. They are obtained from a universal
scaling function calculated within a renormalized scheme for three
particles interacting through pairwise Dirac-delta interaction. The
root-mean-square distance between two atoms of mass $m_A$ in a triatomic
molecule are estimated to be of de order of
${\cal C}\sqrt{\hbar^2/[m_A(E_3-E_2)]}$, where $E_2$ is the dimer and
$E_3$ the trimer binding energies, and ${\cal C}$ is a constant
(varying from $\sim 0.6$ to $\sim 1$) that depends on the ratio between
$E_2$ and $E_3$. Considering previous estimates for the
trimer energies, we also predict the sizes of Rubidium and Sodium
trimers in atomic traps.
\end{abstract}
\maketitle

\section{Introduction}

Weakly bound molecules are large size quantum systems in which the
atoms have an appreciable probability to be found much beyond the
interaction range, and at the same time tiny changes in the
potential parameters can produce huge effects in the properties of
these states\cite{jensen}. The best illustration of such systems
is the experimentally found $^4$He$_2$ dimer\cite{toen}, with
$\sqrt{\langle r^2\rangle}= 52\pm 4\;$\AA \ and binding energy
$E_2$=1.1 + 0.3/ - 0.2 mK\cite{toen00}. Other examples of weakly
bound molecules are found through the experimental realization of
Bose-Einstein condensation (BEC)~\cite{bec95}, where the
possibility to change the effective scattering length of the
low-energy atom-atom interaction in the trap to large positive
values by using an external magnetic field~\cite{bec95,avaria}
can produce very large dimers. In fact, weakly bound molecules in
ultracold atomic traps were reported in Ref.~\cite{donley}. The
binding energy of the $^{87}$Rb$_2$ dimer formed in a
Bose-Einstein condensate was measured with unprecedented
accuracy~\cite{wynar}. Ultracold Na$_2$ molecules have also been
formed through photo-association~\cite{mckenzie}. One should note
that in  limit of an infinite atom-atom scattering length  tuned
by Feshbach resonances in the trap, in principle the Efimov
condition\cite{ef70} can be achieved, in which an infinite
number of weakly-bound trimers exist.
The formation of weakly bound trimers in ultracold atomic systems
have not been reported till now, but recently the recombination
coefficient rate was used to predict trimer binding energies of some
specific atomic species that are being studied in atomic
traps~\cite{recomb}.

Theoretically it is possible to exist weakly bound molecules of
zero-angular momentum states in triatomic systems, as for example,
in the extensively studied $^4$He trimer system (see, e.g. Ref.
\cite{kiev01} and therein). These molecules are special due to the
large spatial size which spreads out much beyond the potential
range\cite{kiev01,roud00}. In such trimer, the calculations of the
mean-square distance of each $^4$He atom to the corresponding
center-of-mass have been performed for the ground and excited
states~\cite{kiev01,roud00}, and also for the mean-square
interatomic distance~\cite{kiev01}.
These sizes are of de order of 5 to 10\ \AA \ for the ground state
of the $^4$He$_3$ molecule, and of about 50 to 90\ \AA \ for the
excited Efimov state~\cite{kiev01}. Therefore, the system heals
through regions that are well outside of the potential range, in
which the wavefunction is essentially a solution of the free
Schr\"odinger equation, and where the physical properties of the
bound system  is defined by few physical scales.
For example, the dimensionless product of the mean-square interatom
distance with the  separation energy of one
atom from the trimer is not far from the unity~\cite{kiev01} in
the ground  and also in the excited  state, despite of the large
difference between such energies $E^{(0)}_3/E^{(1)}_3\approx 50$
($E^{(n)}_3$ is the binding energy of the $n$-th trimer state).
So, as already discussed in Refs.~\cite{ad95,am97,am99,de00}, we
should note that quite naturally the binding energy is the scale
that dominates the physics of the trimer.
One should remember as well that the collapse of the three-body
system in the limit of a zero-range force\cite{th35} makes the
three-body energy one of the scales of the system, beyond the
two-body energy\cite{ad95}.

The calculation of the low-energy properties of the three-body
system can be performed with a renormalization scheme  applied to
three-body equations with $s-$wave zero-range pairwise
potential~\cite{ad95,yama02}. In this approach, one can fix the
three-body ground-state (the three-body physical scale), and the
two-body scattering lengths ~\cite{ad95}. Consequently, all the
detailed information about the short-range force, beyond the
low-energy two-body observables, are retained in only one
three-body physical information in the limit of zero-range
interaction.

In the present work, we first study the mean-square distances of one
atom to the center-of-mass (CM) system and between two atoms in
the ground and excited states of triatomic molecules of the type
$^4{\rm He}_2-$X, where X$\equiv\ ^4$He, $^6$Li, $^7$Li and
$^{23}$Na. Next, using trimer energies derived from the recombination
coefficient rates~\cite{recomb}, we make estimates of the corresponding
sizes of Rubidium and Sodium trimers.
We introduce and calculate scaling functions that describe the different
radii as functions of the physical scales of the triatomic system obtained
in the limit of a zero-range interaction. In this way, we are generalizing
the concept of scaling function, that was previously introduced in
Refs.~\cite{am99} and \cite{yama02} to study the behavior of bound
and excited virtual Efimov states~\cite{ef70} in terms of
triatomic physical scales.

The scaling function depends only on dimensionless ratios of the
binding energies of two and three atoms, and the ratio of masses
of the different atoms.  In that sense our  conclusions apply
equally well to any other low-energy triatomic system. The
validity condition for the scaling relations is that the
interaction range must be small compared to particle distances,
which is the case for weakly bound three-body systems.

The paper is organized as follows. In section II, we present the
Faddeev equations for the spectator functions for a triatomic
system with two equal particles $\alpha$ and a third one $\beta$,
and the form factors from which the different mean-square radii
are obtained. Also in this section we discuss the generalization
of the scaling function defined in Refs.~\cite{am99,yama02} to
describe the different radii.  In section III, we present our
numerical results for  the mean-square distances of one atom in
respect to the CM system and between two atoms in the ground and
excited states of triatomic molecules. Our conclusions are
summarized in section IV.

\section{ Renormalized three-body model and form factors}

In this section, we introduce the generalization of the scaling
function defined in Refs.~\cite{am99} and \cite{yama02}, to be
used to obtain the different radii. We write down the coupled
renormalized equations for the spectator functions and the
expressions for the form factors which allow the calculation of
the different mean-square distances.

\subsection{Subtracted Faddeev Equations}

Throughout this paper we use units such that $\hbar=m_\alpha=1$.
For $\alpha=^4$He, $\hbar^2/m_{^4{\rm He}}=12.12$ K \AA$^2$. After
partial wave projection, the $s-$wave coupled subtracted integral
equations, for two identical particles $\alpha$ and a third one
$\beta$, are given by
\begin{eqnarray}
\chi_{\alpha \alpha}(y)&=&2\tau_{\alpha \alpha}(y;\epsilon_3)
\int_0^\infty dx \frac{x}{y}
G_1(y,x;\epsilon_3)\chi_{\alpha \beta}(x)
\label{chi1} \\
\chi_{\alpha \beta}(y)&=&\tau_{\alpha \beta}(y;\epsilon_3)
\int_0^\infty dx \frac{x}{y}
\left[G_1(x,y;\epsilon_3) \chi_{\alpha \alpha}(x)
\right.  \nonumber \\
 &+& \left. A G_2(y,x;\epsilon_3) \chi_{\alpha \beta}(x)\right] ;
\label{chi2}
\end{eqnarray}
\begin{eqnarray}
\tau_{\alpha \alpha}(y;\epsilon_3)&\equiv &\frac{1}{\pi}
\left[\sqrt{\epsilon_3+\frac{A+2}{4A} y^2} \mp
\sqrt{\epsilon_{\alpha \alpha}} \right]^{-1},  \label{tau1}
\\
\tau_{\alpha \beta}(y;\epsilon_3)&\equiv
&\frac{1}{\pi}\left(\frac{A+1}{2A}\right)^{3/2}
\nonumber \\
&\times& \left[\sqrt{\epsilon_3+\frac{A+2}{2(A+1)} y^2} \mp
\sqrt{\epsilon_{\alpha \beta}}\right]^{-1}\ , \label{tau2}
\\
G_1(y,x;\epsilon_3)&\equiv &\log
\frac{2A(\epsilon_3+x^2+xy)+y^2(A+1)}{2A(\epsilon_3+x^2-xy)+y^2(A+1)}
\nonumber \\
&-& \log\frac{2A(1+x^2+xy)+y^2(A+1)}{2A(1+x^2-xy)+y^2(A+1)} ,
\label{G1} \\
G_2(y,x;\epsilon_3)&\equiv & \log \frac{2(A\epsilon_3
+xy)+(y^2+x^2)(A+1)}{2(A \epsilon_3-xy)+(y^2+x^2)(A+1)} \nonumber
\\ &-& \log \frac{2(A +xy)+(y^2+x^2)(A+1)}{2(A -xy)+(y^2+x^2)(A+1)}.
\label{G2}
\end{eqnarray}
The mass number $A$ is given by the ratio $m_\beta/m_\alpha$. The
plus and minus signs in (\ref{tau1}) and (\ref{tau2}) refer to
virtual and bound two-body subsystems, respectively.

 In the present context that we have three particle systems with
two identical ones, it is worthwhile to call the attention to two
particular definitions of three-body quantum halo states: the 
{\it Borromean} states\cite{zhukov}, where all the two-body subsystems
are virtual ($\alpha-\alpha-\beta)$; and the {\it tango}
states~\cite{robi}, where the $\alpha-\beta$ subsystems is virtual
and $\alpha\alpha$ is bound ($\alpha\alpha-\beta)$. Note that the
virtual pair of particles is denoted with a dash between the
symbols. The Borromean case corresponds to positive signs in front
of the square-root energy of the subsystems in both Eqs.
(\ref{tau1}) and (\ref{tau2}), implying in the weakest attractive
kernel of Eqs.(\ref{chi1}) and (\ref{chi2}) among all the
possibilities of signs in the two-body scattering amplitude. And,
for the tango three-body system, we have negative sign only
in front of $\sqrt{\epsilon_{\alpha\alpha}}$ in Eq.(\ref{tau1}),
with positive sign in front of $\sqrt{\epsilon_{\alpha\beta}}$ in
Eq.(\ref{tau1}). So, a more effective attraction occurs in a 
tango state than in a Borromean case. Of course that, if all the
two-body subsystems are bound, the effective attraction is
maximized; and, if all such subsystems are unbound (virtual), the
effective attraction is minimized.

One can extend the classification scheme of three-body quantum
halo states of the type $\alpha\alpha\beta$~\cite{jensen03},
considering the four possibilities, for increasing values of the
magnitude of the effective attraction in Eqs. (\ref{chi1}) and
(\ref{chi2}). The weakest attractive situation corresponds to the
previous defined Borromean-type (only virtual subsystems)
($\alpha-\alpha-\beta$). The tango situation
($\alpha\alpha-\beta$) is followed by an three-body system with
$\alpha-\alpha$ virtual and $\alpha\beta$ bound, that we represent
by ($\alpha\beta\alpha$) halo system. Three-body system with the
strongest effective attraction has all the subsystems bound and it
is represented by ($\alpha\alpha\beta$).

We solve Eqs.~(\ref{chi1}-\ref{G2}) in units such that the
three-body subtraction point $\mu_{(3)}$ is equal to 
one\cite{yama02}. The corresponding dimensionless quantities
are: $\epsilon_3\equiv E_3/\mu_{(3)}^2,$ $\epsilon_{\alpha \alpha} \equiv
E_{\alpha \alpha} /\mu^2_{(3)},$ $\epsilon_{\alpha \beta} \equiv
E_{\alpha \beta}/\mu^2_{(3)}.$ The three-body physical quantities can be
written in terms of the three-body binding energy $E_3$ when first
the value of $\mu^2_{(3)}$ is determined from the known value of
$E_3$. Therefore, the results for the renormalized model appear
when the subtraction point energy is written as a function of
$E_3$ and consequently the three-body quantities naturally scale
with $E_3$. Finally, the scaling functions are obtained when the
dimensionless product of physical quantities are written as a
function of the ratios between two-body energies and $E_3$.

\subsection{Scaling functions for the radii}

The existence of a three-body scale implies in the low energy
universality found in three-body systems, or correlations between
three-body observables~\cite{fre87a,ad95}. In the scaling
limit\cite{am97,yama02}, one has
\begin{eqnarray}
&&{\cal{O}}\left(E,
E_{3},E_{\alpha\alpha},E_{\alpha\beta},\right)(E_{3})^{-\eta} =
\nonumber \\
&&{\cal A}\left(\sqrt{E/E_{3}},\sqrt{E_{\alpha\alpha}/E_{3}},
,\sqrt{E_{\alpha\beta}/E_{3}}, A\right) \ , \label{o}
\end{eqnarray}
where $\cal O$ is a general observable of the three-body system at
energy $E$, with dimension of energy to the power $\eta$.   In the
present paper we discuss only the situation that we have only
bound subsystems $(\alpha\alpha\beta)$; however, the analysis
could be easily extended to other three-body halo systems, as the
Borromean, tango and $(\alpha\beta\alpha)$ systems.

In the case of the mean-square separation distances, $\langle
r^2_\gamma\rangle$ with $\gamma=\ \alpha$ or $\beta$, i.e, the
distance of the atom $\gamma$ to the CM; and $\langle
r^2_{\alpha\gamma}\rangle$, i.e, the distance between the atoms
$\alpha$ and $\gamma$, the scaling functions are of the form:
\begin{eqnarray}
\sqrt{\langle r^2_{\gamma}\rangle S_{3}} = {\cal R}_\gamma\left(
\sqrt{\epsilon_{\alpha\alpha}/\epsilon_{3}},
,\sqrt{\epsilon_{\alpha\beta}/\epsilon_{3}}, A\right) \ ,
\label{rg}
\end{eqnarray}
and
\begin{eqnarray}
\sqrt{\langle r^2_{\alpha\gamma}\rangle S_{3}} = {\cal
R}_{\alpha\gamma}\left(
\sqrt{\epsilon_{\alpha\alpha}/\epsilon_{3}},
,\sqrt{\epsilon_{\alpha\beta}/\epsilon_{3}}, A\right) \ ,
\label{rag}
\end{eqnarray}
where $S_3$ is the smallest separation energy of the three-body
system, i.e.,
$S_3=min\left(E_3-E_{\alpha\alpha},E_3-E_{\alpha\beta}\right)$.
 Two particular situations are worthwhile mentioning, one is
the case of trimer systems ($A=1$), where the scaling functions
above reduce to:
\begin{eqnarray}
\sqrt{\langle r^2_{\gamma}\rangle S_{3}} = {\cal R}_\gamma\left(
\sqrt{\epsilon_{2}/\epsilon_{3}}\right) \ , \label{rg1}
\end{eqnarray}
and
\begin{eqnarray}
\sqrt{\langle r^2_{\alpha\alpha}\rangle S_{3}} = {\cal
R}_{\alpha\gamma}\left( \sqrt{\epsilon_{2}/\epsilon_{3}}\right) \
. \label{rag1}
\end{eqnarray}
The other special situation is found for
$\epsilon_{\alpha\gamma}=0$ where the dimensionless product of the
square radii and triatomic binding energy depend only on the mass
ratio:
\begin{eqnarray}
\sqrt{\langle r^2_{\gamma}\rangle E_{3}} = {\cal
R}_\gamma\left(A\right) \ , \label{rg2}
\end{eqnarray}
and
\begin{eqnarray}
\sqrt{\langle r^2_{\alpha\gamma}\rangle E_{3}} = {\cal
R}_{\alpha\gamma}\left(A\right) \ . \label{rag2}
\end{eqnarray}

\subsection{Form factors}

The mean-square radii are calculated from the derivative of the
Fourier transform of the respective matter density in respect to
the square of the momentum transfer. The Fourier transform of the
one and two-body densities define the respective form factors,
$F_\beta(q^2)$ and $F_{\alpha\gamma}(q^2)$, as a
function of the dimensionless momentum transfer $\vec q$.
For the mean-square radius of the particle $\gamma$ ($=\alpha$ or
$\beta$) to CM, we have
\begin{eqnarray}
\langle r^2_\gamma\rangle=
-6\left(1-\frac{m_\gamma}{2m_\alpha+m_\beta}\right)^2
\frac{dF_\gamma(q^2)}{dq^2}\bigg|_{q^2=0}, \label{ra}
\end{eqnarray}
where
\begin{eqnarray}
F_\alpha( q^2)&=&\int d^3yd^3z
\Psi_{\alpha\beta}(\vec{y}+\frac{\vec{q}}{2},\vec{z})
\Psi_{\alpha\beta}(\vec{y}-\frac{\vec{q}}{2},\vec{z}) \nonumber\\
F_\beta( q^2)&=&\int d^3yd^3z
\Psi_{\alpha\alpha}(\vec{y}+\frac{\vec{q}}{2},\vec{z})
\Psi_{\alpha\alpha}(\vec{y}-\frac{\vec{q}}{2},\vec{z}).
\label{F3}
\end{eqnarray}
And, for the mean-square distance between the particles $\alpha$
and $\gamma$, we have
\begin{eqnarray}
\langle r^2_{\alpha\gamma}\rangle= -6 \frac{dF_{\alpha\gamma}(
q^2)}{dq^2}\bigg|_{q^2=0}, \label{rab}
\end{eqnarray}
where
\begin{eqnarray}
F_{\alpha\gamma}( q^2)=\int
d^3yd^3z\Psi_{\alpha\gamma}(\vec{y},\vec{z}+\frac{\vec q}{2})
\Psi_{\alpha\gamma}(\vec{y},\vec{z}-\frac{\vec{q}}{2})\ .
\label{F4}
\end{eqnarray}
The above triatomic wave-functions in momentum space are given in
terms of the spectator functions $\chi_{\alpha\gamma}$:
{\small
\begin{eqnarray}
\nonumber
&&\Psi_{\alpha\alpha}(\vec y,\vec z)=\left(\frac{1}{\epsilon_3+
\frac{A+2}{4A} \vec y^2+\vec z^2}-\frac{1}{1+
\frac{A+2}{4A} \vec y^2+\vec z^2}\right) \\ \nonumber
&&\times\left(\chi_{\alpha\alpha}(|\vec y|)+
\chi_{\alpha\beta}(|\vec z-\frac{\vec y}{2}|)+ \
\chi_{\alpha\beta}(|\vec z+\frac{\vec y}{2}|)\right),
\\
\label{psi}\\ \nonumber
&&\Psi_{\alpha\beta}(\vec y,\vec z)= \\ \nonumber
&&\left(\frac{1}{\epsilon_3+\frac{A+1}{2A} \vec z^2+
\frac{A+2}{2(A+1)}\vec y^2}-\frac{1}
{1+\frac{A+1}{2A} \vec z^2+\frac{A+2}{2(A+1)}
\vec y^2}\right) \\ \nonumber
&&\times\left(\chi_{\alpha\alpha}(|\vec z -
\frac{A \vec y}{A+1}|)+\chi_{\alpha\beta}(|\vec y|)
+ \chi_{\alpha\beta}(|\vec z+\frac{\vec y}{A+1}|)\right),
\end{eqnarray}
}
where $\vec z$ is the relative momentum of the pair
and $\vec y$ is the relative momentum of the spectator particle
to the pair in units of $\mu_{(3)}=1$. Note that the sub-indices
of $\Psi$ in Eq.(\ref{psi}) just denote the pair of Jacobi
relative momenta used to evaluate the wave-function.  For
$\alpha\gamma$ with $\gamma\ = \ \alpha $ or $\beta $, one has the
relative momentum between $\alpha$ and $\gamma$ and the relative
momentum of the third particle to the center-of-mass of the
system $\alpha \gamma$.

\section{Results for triatomic radii}

Our analysis has considered some particular three-body molecular
systems, in which the three-body ground-state energy and the
corresponding energies of the two-body subsystem is known
theoretically for $^4$He trimer\cite{kiev01}, $^4$He$_2-^6$Li,
$^4$He$_2-^7$Li and $^4$He$_2-^{23}$Na \cite{Yuan}. In
Ref.~\cite{kiev01}, the authors have considered realistic two-body
interactions; their results for the ground and excited states
radii are appropriate for our purpose of comparing with the
present scaling approach.

The ground and excited Efimov state energies of the $^4$He$_3$
molecule were extensively studied in the scaling approach of Refs.
\cite{am99,yama02} with results that are in very good agreement
with realistic calculations. This lead us to conclude that other
details (beyond the dimer and trimer ground-state energies)
presented in the realistic interactions, that have been used, are
quite irrelevant to the existence of Efimov states. These features
validates a universal scaling function, relating the trimer
ground-state, the dimer and the weakly bound excited three-body
energy state. Realistic calculations for the excited states of
$^4$He trimer approaches reasonably well the scaling
limit\cite{am99,yama02}, which suggests to investigate  the
scaling limit of other observables like the different radii
defined in Eqs. (\ref{ra}) and (\ref{rab}). The conditions for the
validity of the present approach are that the atoms should have a
very shallow and short-ranged two-body interaction and the binding
energy close to zero, i.e., the ratio between the interaction
range and dimer size should be much smaller than 1. These are
indeed the cases we are considering.

The results for the radii of $^4$He$_3$ molecule in the ground and
excited state are shown in Fig. 1, in the form of a scaling plot.
The dimensionless products $\sqrt{\langle r^2_{\alpha}\rangle
S_3}$  and  $\sqrt{\langle r^2_{\alpha\alpha}\rangle S_3}$  as
functions of $\sqrt{E_2/E_3}$ are shown in the figure and compared
to  the realistic calculations, obtained from Ref.\cite{kiev01}.
Our calculations for the ground and excited state are practically
the same, which would be the case if the energies in respect
to $\mu^2$ are in fact going to zero, i.e., the scaling limit. The
results for $\sqrt{\langle r^2_{\alpha}\rangle S_3}$ and
$\sqrt{\langle r^2_{\alpha\alpha}\rangle S_3}$ for the excited
state are in good agreement with the realistic result. However,
for the ground state the results show a deviation of about 20\%.
The excited state size is about ten times larger than the
corresponding size of the ground state. Therefore, the scaling
limit is better approached in the excited state, which is much
larger than the interaction range, which is not strictly valid for
the ground state, and consequently deviations in the scaling plot
are stronger for this state.

\begin{figure}[thbp]
\setlength{\epsfxsize}{0.9\hsize}
\setlength{\epsfysize}{0.3\vsize}
\centerline{\epsfbox{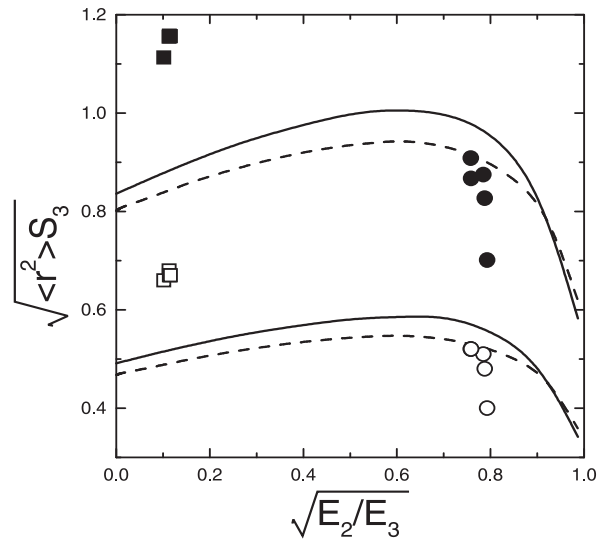}}
\caption[dummy0]{The dimensionless products $\sqrt{\langle
r^2_{\alpha}\rangle S_3}$ (lower curves) and  $\sqrt{\langle
r^2_{\alpha\alpha}\rangle S_3}$ (upper curves)  as functions of
$\sqrt{E_2/E_3}$. Our results for the ground state and first
excited state are shown respectively, by solid and dashed lines.
Realistic calculations from Ref.{\protect{\cite{kiev01}}}, for
$\sqrt{\langle r^2_{\alpha}\rangle S_3}$ are given by empty
squares (ground state) and empty circles (excited state); and, for
$\sqrt{\langle r^2_{\alpha\alpha}\rangle S_3}$, by full squares
(ground state) and full circles (excited state). }\label{fig1}
\end{figure}

\begin{figure}[thbp]
\setlength{\epsfxsize}{0.9\hsize}
\setlength{\epsfysize}{0.3\vsize}
\centerline{\epsfbox{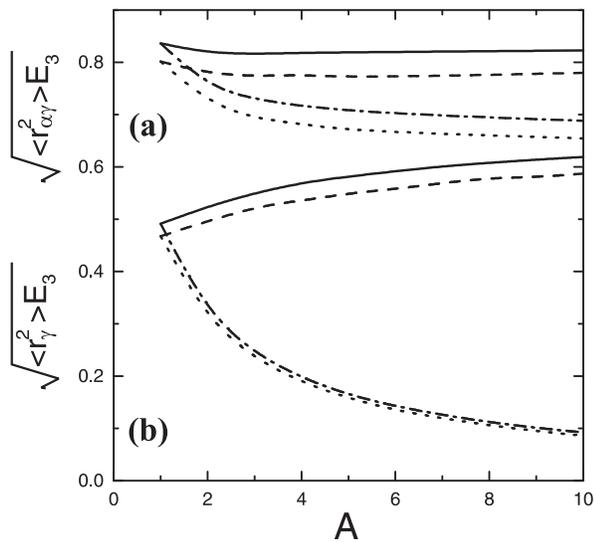}}
\caption[dummy0]{
For the triatomic $\alpha\alpha\beta$ system,
with $\gamma\equiv $ $\alpha$, $\beta$,
it is shown the dimensionless products $\sqrt{\langle
r^2_{\alpha\gamma}\rangle E_3}$ [upper (a) plots]
and $\sqrt{\langle r^2_{\gamma}\rangle E_3}$ [lower (b) plots],
as functions of $A\equiv m_\beta/m_\alpha$, in the
limit $E_{\alpha\alpha}=E_{\alpha\beta}=0$.
$\sqrt{\langle r^2_{\gamma}\rangle}$ is the root-mean-square
distance of particle $\gamma$ from the center-of-mass,
and $\sqrt{\langle r^2_{\alpha\gamma}\rangle}$ is the root-mean-square
distance between the particles $\alpha$ and $\gamma$.
The results for the ground-state ($N=0$) are shown with solid line
($\gamma=\alpha$) and dot-dashed line ($\gamma=\beta$);
and, for the excited state ($N=1$), with dashed line ($\gamma=\alpha$)
and dotted line ($\gamma=\beta$).
} \label{fig3}
\end{figure}

In Fig. 2, the results for the dimensionless products
$\sqrt{\langle r^2_{\alpha\alpha}\rangle E_3}$, $\sqrt{\langle
r^2_{\alpha\beta}\rangle E_3}$, and $\sqrt{\langle
r^2_{\gamma}\rangle E_3}$, as functions of $A= m_\beta/m_\alpha$,
for $E_{\alpha\alpha}=E_{\alpha\beta}=0$ are shown. We perform
calculations for the ground ($N=0$) and excited ($N=1$) states, as
indicated in the figure. One observe in the upper frame of Fig. 2,
that the results almost saturates above $A\approx 3$ to the values
found in the limit of $A=\infty$. The calculations for $A=\infty$
give for $\langle r^2_{\alpha\alpha}\rangle$ the values of $0.69/
E_3$ for N=0 and $0.61/E_3$ for $N=1$. Therefore, for the ground states, 
the root mean-square distance between two $^4$He in the triatomic molecules 
can be estimated by 0.83$\sqrt{\hbar^2/(E_3m_{^4{\rm He}})}$, in the limit
of zero pairwise binding energies. Our results for $\langle
r^2_{\alpha\beta}\rangle$ are $0.45/E_3$ for $N=0$ and
$0.40/E_3$ for $N=1$. The saturation value for $\sqrt{\langle
r^2_{\alpha\alpha}\rangle E_3}$ is achieved fast with increasing
$A$ than for $\sqrt{\langle r^2_{\alpha\beta}\rangle E_3}$ which
depends on the difference in the masses of the atomic pair.
The mean-square distance of one of the atoms $\gamma$ ($= \alpha$ or
$\beta$) to the center-of-mass of the molecule can be obtained from the
lower frame of Fig. 2, where we plot $\sqrt{\langle r^2_{\gamma}\rangle
E_3}$ as a function of $A$.
One sees that, for the infinitely heavy $\beta$-atom, the results for
$\langle r^2_{\alpha}\rangle$ are the same of the $\langle
r^2_{\alpha\beta}\rangle$, while
$\sqrt{\langle r^2_{\beta}\rangle E_3}\ = \ 0$ as the heavy
particle should rest in the CM of the molecule in this limit.
We remind the reader that the ratio between the binding energies
for $N=0$ and $N=1$ is about 500 for $A=1$ \cite{ef70}
and 300 for $A=10$, while the dimensionless products of square
radius and energy changes only around 10\%.

In the above, we have considered examples of molecules with two
helium atoms. However, our results presented in Figs. 1 and 2 are more
general, such that we can extend the estimates to other atomic systems.
Of particular interest is the analysis of possible formation of
molecular systems in experiments with ultracold trapped gases.
By considering, for example, the estimates of trimer energies
obtained from the recombination coefficient, given in
Ref.~\cite{recomb}, within our approach we can predict the
corresponding trimer sizes. In this case, we have $\alpha=\beta$
and $A=1$ in the previous equations and in the figures. However,
our unit for a specific trimer of an atom with $A$ nucleons will
be $\hbar^2/m_{A}=(48.48/A)$ K \AA$^2$.

In Table I, we present our results for the different radii of the
ground state ($N=0$) of the weakly bound molecular systems
$\alpha\alpha\beta$, where $\alpha\equiv ^4$He and $\beta=$
$^4$He, $^{6}$Li, $^{7}$Li and $^{23}$Na, obtained from the known
theoretical values of $E_3^{(0)}$, $E_{\alpha\alpha}$ and
$E_{\alpha\beta}$\cite{Yuan}. Our calculation for $^4$He$_3$ of
$\sqrt{\langle r^2_{\alpha\alpha}\rangle}$ gives 9.45 \AA \
which is 14\% off the value 11\AA \ obtained in the realistic
variational calculations of Ref.~\cite{kiev01}. The same
quality of agreement is found for $\sqrt{\langle r^2_{\alpha}\rangle}$
which in our calculation is 5.55\AA \ compared to 6.4\AA \ of
Ref.~\cite{kiev01}. The quality of the reproduction of the realistic
results by our calculations are quite surprising in view of the
simplicity of the present approach, where the only physical inputs
are the values of the dimer and trimer binding energies. The several
different radii of the molecules $^4$He$_2$-$^{6}$Li,
$^4$He$_2$-$^{7}$Li and $^4$He$_2$-$^{23}$Na have values larger
than those found in the $^4$He$_3$ which makes plausible that our
predictions are even better in quality.

\begin{table}
\caption[dummy0] {Results for different radii of the molecular
systems $\alpha\alpha\beta$, where $\alpha\equiv ^4$He and
$\beta$ is identified in the first column. The ground-state
energies of the triatomic molecules and the corresponding energies
of the diatomic subsystems, obtained from Ref.~\cite{Yuan}, are given in
the second, third and forth columns. $\langle
r^2_{\alpha\gamma}\rangle$ is the corresponding mean-square
distance between the particles $\alpha$ and $\gamma$ ($= \alpha,
\beta$). $\langle r^2_{\gamma}\rangle$ is the mean-square distance
of $\gamma$ from the trimer center-of-mass. } \vskip 0.5cm
\begin{tabular}{|c|c|c|c|c|c|c|c|} \hline
$\beta$ & $E_3^{(0)}$ & $E_{\alpha\alpha}$ & $E_{\alpha\beta}$ &
$\sqrt{\langle r^2_{\alpha\alpha}\rangle}$ & $\sqrt{\langle
r^2_{\alpha\beta}\rangle} $  & $\sqrt{\langle r^2_\alpha\rangle}$ &
$\sqrt{\langle r^2_\beta\rangle}$ \\
&(mK)&(mK)&(mK)& (\AA) & (\AA)&(\AA)&(\AA)\\
\hline \hline
$^4$He    & 106.0& 1.31 & 1.31 & 9.45 & 9.45 & 5.55 & 5.55  \\
$^6$Li    & 31.4 & 1.31 & 0.12 &16.91 &16.38 & 10.50& 8.14  \\
$^7$Li    & 45.7 & 1.31 & 2.16 &14.94 &13.88 & 9.34 & 6.31  \\
$^{23}$Na &103.1 & 1.31 &28.98 &11.66 & 9.54 & 8.12 & 1.94  \\
\hline
\end{tabular}
\end{table}

\begin{table}
\caption[dummy0] {Results for the size of trimer systems
predicted in Ref.\cite{recomb}. The sizes are given by
$\sqrt{\langle r^2_{\alpha}\rangle}$,
the root-mean-square distance between the atom $\alpha$
and the center-of-mass of the trimer system.
The atoms of the trimer are identified in the first column.
For each dimer energy, given in the second column, we have
two possible trimer energies (columns 3 and 5), with
the corresponding radii given in the columns 4 and 6.
The trimer estimates, given in Ref.~\cite{recomb}, for
$^{87}$Rb$|1,-1\rangle$, are for noncondensed ($^*$) and
condensed ($^\dagger$) trapped atoms.}
\vskip 0.5cm
\begin{tabular}{|c|c|c|c|c|c|} \hline
$Atom$          &
$E_2$           &
$E_3$           &
$\sqrt{\langle r^2_{\alpha}\rangle} $ &
$E_3^\prime$    &
$\sqrt{\langle r^2_{\alpha}\rangle^\prime}$ \\
&(mK)&(mK)&(\AA)&(mK)&(\AA)\\
\hline \hline
$^{23}$Na$|1,-1\rangle$
& 2.85& 7.75 & 12 & 3.06 & 38 \\
$^{87}$Rb$|1,-1\rangle^*$
& 0.17& 0.56 & 22 & 0.175 & 114 \\
$^{87}$Rb$|1,-1\rangle^\dagger$
& 0.17& 0.47 & 25 & 0.183 & 91 \\
$^{85}$Rb$|2,-2\rangle$
& $1.3\times 10^{-4}$ & $2.4\times10^{-4}$ & 1293 &
$1.7\times 10^{-4}$  & 1944 \\
\hline
\end{tabular}
\end{table}

We point out that the results in Table I, for the $^4$He dimer
sizes inside the molecules shrink in respect to the free value of
52\AA, due to the large values of the trimer binding energies.
Qualitatively this is explained just by considering that the
dimer size scales roughly with the inverse of the square-root of
its binding energy inside the molecule, which can be estimated to
be 2/3 of the molecule binding, from which one finds for that the
dimer has sizes around 10\AA, close to the ones we have found in
Table I.

In Table II, we are also presenting results for different radii of
the trimers predicted in Ref. \cite{recomb}, from where we obtain
the energy of the dimer and the most weakly bound trimer energies
of $^{23}$Na$|F=1,m_F=-1\rangle$, $^{87}$Rb$|F=1,m_F=-1\rangle$,
and $^{85}$Rb$|F=2,m_F=-2\rangle$, where $|F,m_F\rangle$ is the
respective hyperfine states of the total spin $F$. We are
presenting the mean-square distance from each atom to the
center-of-mass of the corresponding trimer. From Fig. 1, one can
also obtain the corresponding mean-square distance between the
atoms. We observe that one value of the recombination rate is
consistent with two values of the most weakly bound trimer energy,
as discussed in Ref.~\cite{recomb}. Therefore, we present two
possible values for the radii that are consistent with the
corresponding weakly bound trimer energies. Actually, we should
also mention that, in a trap, one can achieve dimer and trimer
molecules with very large sizes, following the possibility to
alter the corresponding two-body scattering length~\cite{avaria}.

Finally, it is interesting to recall the results for the
hyperradius calculations 
obtained by Jensen and collaborators \cite{jensen03}. 
From their scaling plot, one can observe that the hyperradius of a
tango system is bigger than the hyperradius of a Borromean system,
for the same three-body energy. 
This point can become very clear, for instance, if we take as
an example the results shown for the $^3_\Lambda H$ (filled 
circles in Fig.2 of Ref.\cite{jensen03}), and
estimate the dimensionless product of observables, 
$\langle\rho^2\rangle mB/\hbar^2$ (product of the x-axis and 
y-axis in Fig.2 of Ref.\cite{jensen03}, where
$\rho$ is the hyperradius and $B$ the three-body binding energy). 
When going from the tango to the Borromean configuration, this
product will decrease.
Therefore, if one keeps the same binding energy,
the Borromean system would be more compact than the tango system.
Extending this analysis to ($\alpha\beta\alpha$) halos, where
$(\alpha-\alpha)$ is virtual and ($\alpha\beta$) is bound,
and also to all-bound pairs ($\alpha\alpha\beta$), one should
expect that the sizes increase when going from Borromean states to
halos with all-bound subsystems, while keeping the three-body
energy fixed. 
Within our approach, the scaling relations are expected to be 
followed  in all the cases, in the limit of a zero-range interaction. 
The Borromean halo (the less attractive system), in order to have the
same three-body energy as a tango state, should be more compact, in
agreement with the scaling plot of Ref.~\cite{jensen03}. 
In a realistic case, our scheme is expect to produce better results in
the all-bound case, when comparing systems with the same three-body
energies. This occurs because the all-bound case would have the most
extended wave-function; consequently, the range of the potential, in
relation to the size, would have the smallest value, satisfying
better the validity condition for the scaling relations, which is
that the interaction range must be small compared to particle
distances.  

\section{Conclusions}

The mean-square radii of the triatomic molecules, $^4$He$_3$,
$^4$He$_2$-$^{6}$Li, $^4$He$_2$-$^{7}$Li and $^4$He$_2$-$^{23}$Na
are calculated using a renormalized three-body model with a
pairwise Dirac-delta interaction, having as physical inputs only
the values of the binding energies of the diatomic and triatomic
molecules. Presently, we have considered molecular three-body systems
with bound subsystems, due to the available data. 
The renormalized zero range model can also be 
applied to the cases where at least one of the subsystems is virtual. 
When comparing systems with the same three-body energies and potential
ranges, he validity of the renormalized
zero range model is expected to be better in the all-bound case, because
this case corresponds to the most extended wave-function.

The validity of the present framework is substantiated
by the agreement of our results for the different radii with the
realistic potential model calculations of Ref.\cite{kiev01} for
$^4$He$_3$ ground and excited states which are within about 14\%.
These results are quite surprising in view of the simplicity of
the approach, where the only physical inputs are the values of diatomic
and triatomic binding energies. We predicted for the first time, as far
we know, the values of several different radii for $^4$He$_2$-$^{6}$Li,
$^4$He$_2$-$^{7}$Li and $^4$He$_2$-$^{23}$Na molecules, from the
theoretical values of the binding energies calculated in
Ref.~\cite{Yuan}. These other molecules are in general larger than the
$^4$He-trimer indicating that our radii predictions for these triatomic
ground states can be even better in quality than those found for
$^4$He$_3$.

In view of the actual relevance of ultracold atomic systems that
are being experimentally studied, and the possibility to observe
the formation of molecular systems in trapped condensates, we also
present results for the sizes of rubidium and sodium trimers,
considering the binding energies that were recently
estimated~\cite{recomb} from analysis of the corresponding
three-body recombination coefficients.

We would like to thank Funda\c c\~ao de Amparo \`a Pesquisa do
Estado de S\~ao Paulo (FAPESP) for partial support. LT and TF also
thank partial support from Conselho Nacional de Desenvolvimento
Cient\'{\i}fico e Tecnol\'ogico (CNPq).


\begin{thebibliography}{}

\bibitem{jensen}E. Nielsen, D. V. Fedorov, A. S.  Jensen, E. Garrido,
Phys. Rep. {\bf 347}, 373 (2001).

\bibitem{toen} W. Sch\"ollkopf and J. P. Toennies, Science {\bf
266}, 1345 (1994); K.T. Tang, J.P. Toennies, and C.L. Liu, Phys.
Rev. Lett. {\bf 74}, 1546 (1995).

\bibitem{toen00} R.E. Grisenti, W. Sch\"ollkopf, J.P. Toennies,
G.C. Hegerfeldt, T. K\"ohler, and M. Stoll; Phys. Rev. Lett. {\bf
85}, 2284 (2000).

\bibitem{bec95}
M.H. Anderson, J.R. Ensher, M.R. Matthews, C.E. Wieman, E.A.
Cornell, Science {\bf 269}, 198 (1995); C.C. Bradley, C.A.
Sackett, J.J. Tollet, and R.G. Hulet, Phys. Rev. Lett. {\bf 75},
1687 (1995); K.B. Davis, M.-O. Mewes, M.R. Andrews, N.J. van
Druten, D.S. Durfee, D.M. Kurn, W. Ketterle, Phys. Rev. Lett. {\bf
75}, 3969 (1995).

\bibitem{avaria} S. Inouye, M.R. Andrews, J. Stenger, H.-J. Miesner,
D.M. Stamper-Kurn, W. Ketterle, Nature {\bf 392}, 151 (1998); E.
Timmermans, P. Tommasini, M. Hussein, and A. Kerman, Phys. Rep.
{\bf 315}, 199 (1999).

\bibitem{donley}
N.R. Claussen, E.A. Donley, S.T. Thompson, and C. E. Wieman,
Phys. Rev. Lett. {\bf 89}, 010401(2002);
E.A. Donley, N.R. Claussen, S.T. Thompson, and C. E. Wieman,
Nature (London) {\bf 417}, 529 (2002).

\bibitem{wynar} R.H. Wynar, R.S. Freeland, D.J. Han, C. Ryu,
and D.J. Heinzen,  Science {\bf 287}, 1016 (2000).

\bibitem{mckenzie} C. Mckenzie {\it et al.},
Phys. Rev. Lett. {\bf 88}, 120403 (2001).

\bibitem{ef70} V. Efimov, Phys. Lett. {\bf B 33}, 563 (1970);
Nucl. Phys. {\bf A362}, 45 (1981);
V. Efimov, Comm. Nucl. Part. Phys. {\bf 19}, 271 (1990)
and references therein.

\bibitem{recomb} M.T. Yamashita, T. Frederico, A. Delfino, L.
Tomio, cond-mat/0206317, submitted to Phys. Rev. A.

\bibitem{kiev01} P. Barletta and A. Kievsky, Phys. Rev. {\bf A64},
042514 (2001).

\bibitem{roud00} V. Roudnev, S. Yakovlev, Chem. Phys. Lett. {\bf
328}, 97 (2000).

\bibitem{ad95} S. K. Adhikari, T.Frederico and I.D. Goldman,
Phys. Rev. Lett. {\bf 74}, 487 (1995); S.K. Adhikari and T.
Frederico, Phys. Rev. Lett. {\bf 74}, 4572(1995).

\bibitem{am97} A.E.A. Amorim, L. Tomio and T. Frederico,
Phys. Rev. C{\bf 56}, R2378 (1997).

\bibitem{am99} T. Frederico, A. Delfino, A.E.A. Amorim and L. Tomio,
Phys. Rev. A{\bf 60}, R9 (1999).

\bibitem{de00} A. Delfino, T. Frederico, L. Tomio,
Few-Body Syst. {\bf 28}, 259 (2000); J. Chem. Phys. {\bf 113},
7874 (2000).

\bibitem{th35} L.H. Thomas, Phys. Rev. {\bf 47}, 903 (1935).

\bibitem{yama02} M.T. Yamashita, T. Frederico, A. Delfino, L.
Tomio, Phys. Rev. {\bf A66}, 052702 (2002).

\bibitem{zhukov} M.V. Zhukov, B.V. Danilin, D.V. Fedorov, J.M. Bang,
I.J. Thompson and J. Vaagen, Phys. Rep. {\bf 231}, 151 (1993).

\bibitem{robi} F. Robichaux, Phys. Rev. {\bf A60}, 1706 (1999).

\bibitem{jensen03} A.S. Jensen, K. Riisager, D.V. Fedorov and
E. Garrido, Europhys. Lett. {\bf 61}, 320 (2003).

\bibitem{fre87a}T. Frederico, I. D. Goldman, Phys. Rev. C{\bf 36}
R1661 (1987).

\bibitem{Yuan} J. Yuan and C.D. Lin, J. Phys. {\bf B31}, L637
(1998).
\end{thebibliography}
\end{document}